\documentclass[aps,prb,twocolumn,superscriptaddress,showpacs,amsmath,amssymb]{revtex4}
\usepackage{CJK}

\usepackage{}
\usepackage{mathrsfs}
\usepackage{amssymb}
\usepackage{txfonts}
\usepackage{tipa}

\usepackage{graphicx}
\usepackage{dcolumn}
\usepackage{bm}

\begin{document}
\begin{CJK*}{GB}{} 

\title{Spin-Wave and Electromagnon Dispersions in Multiferroic MnWO$_4$ as Observed by Neutron Spectroscopy: Isotropic Heisenberg Exchange versus Anisotropic Dzyaloshinskii-Moriya Interaction}
\CJKfamily{gbsn}
\author{Y. Xiao}
\email[y.xiao@fz-juelich.de]{}
\affiliation{J\"{u}lich Centre for Neutron Science JCNS and Peter Gr\"{u}nberg Institut PGI, JARA-FIT, Forschungszentrum J\"{u}lich GmbH, 52425 J\"{u}lich, Germany}

\author{C. M. N. Kumar}
\email[n.kumar@fz-juelich.de]{}
\affiliation{J\"{u}lich Centre for Neutron Science JCNS, Forschungszentrum J\"{u}lich GmbH, Outstation at SNS, POB 2008, 1 Bethel Valley Rd. Oak Ridge, TN 37831-6473, USA}
\affiliation{Chemical and Engineering Materials Division, Oak Ridge National Laboratory, Oak Ridge, Tennessee 37831, USA}

\author{S. Nandi}
\affiliation{J\"{u}lich Centre for Neutron Science JCNS and Peter Gr\"{u}nberg Institut PGI, JARA-FIT, Forschungszentrum J\"{u}lich GmbH, 52425 J\"{u}lich, Germany}
\affiliation{J\"{u}lich Centre for Neutron Science JCNS at Heinz Maier-Leibnitz Zentrum, Forschungszentrum J\"{u}lich GmbH, Lichtenbergstra{\ss}e 1, 85747 Garching, Germany}
\affiliation{Department of Physics, Indian Institute of Technology, Kanpur 208016, India}

\author{Y. Su}
\affiliation{J\"{u}lich Centre for Neutron Science JCNS at Heinz Maier-Leibnitz Zentrum, Forschungszentrum J\"{u}lich GmbH, Lichtenbergstra{\ss}e 1, 85747 Garching, Germany}

\author{W. T. Jin}
\affiliation{J\"{u}lich Centre for Neutron Science JCNS and Peter Gr\"{u}nberg Institut PGI, JARA-FIT, Forschungszentrum J\"{u}lich GmbH, 52425 J\"{u}lich, Germany}
\affiliation{J\"{u}lich Centre for Neutron Science JCNS at Heinz Maier-Leibnitz Zentrum, Forschungszentrum J\"{u}lich GmbH, Lichtenbergstra{\ss}e 1, 85747 Garching, Germany}

\author{Z. Fu}
\affiliation{J\"{u}lich Centre for Neutron Science JCNS at Heinz Maier-Leibnitz Zentrum, Forschungszentrum J\"{u}lich GmbH, Lichtenbergstra{\ss}e 1, 85747 Garching, Germany}

\author{E. Faulhaber}
\affiliation{Heinz Maier-Leibnitz Zentrum (MLZ), Technische Universit\"{a}t M\"{u}nchen, Lichtenbergstra{\ss}e 1, 85748 Garching, Germany}
\affiliation{Helmholtz-Zentrum Berlin f\"{u}r Materialien und Energie, Hahn-Meitner-Platz 1, D-14109 Berlin, Germany}

\author{A. Schneidewind}
\affiliation{J\"{u}lich Centre for Neutron Science JCNS at Heinz Maier-Leibnitz Zentrum, Forschungszentrum J\"{u}lich GmbH, Lichtenbergstra{\ss}e 1, 85747 Garching, Germany}
\affiliation{Helmholtz-Zentrum Berlin f\"{u}r Materialien und Energie, Hahn-Meitner-Platz 1, D-14109 Berlin, Germany}

\author{Th. Br\"{u}ckel}
\affiliation{J\"{u}lich Centre for Neutron Science JCNS and Peter Gr\"{u}nberg Institut PGI, JARA-FIT, Forschungszentrum J\"{u}lich GmbH, 52425 J\"{u}lich, Germany}
\affiliation{J\"{u}lich Centre for Neutron Science JCNS at Heinz Maier-Leibnitz Zentrum, Forschungszentrum J\"{u}lich GmbH, Lichtenbergstra{\ss}e 1, 85747 Garching, Germany}

\date{\today}

\begin{abstract}

High resolution inelastic neutron scattering reveals that the elementary magnetic excitations in multiferroic MnWO$_4$ consist of low energy dispersive electromagnons in addition to the well-known spin-wave excitations. The latter can well be modeled by a Heisenberg Hamiltonian with magnetic exchange coupling extending to the 12$^{th}$ nearest neighbor. They exhibit a spin-wave gap of 0.61(1) meV. Two electromagnon branches appear at lower energies of 0.07(1) meV and 0.45(1) meV at the zone center. They reflect the dynamic magnetoelectric coupling and persist in both, the collinear magnetic and paraelectric AF1 phase, and the spin spiral ferroelectric AF2 phase. These excitations are associated with the Dzyaloshinskii-Moriya exchange interaction, which is significant due to the rather large spin-orbit coupling.

\end{abstract}

\pacs{75.85.+t, 78.70.Nx, 75.30.Ds, 75.40.Gb}
\maketitle
\end{CJK*}

\section{\label{sec:level1} Introduction}

Multiferroics with strong coupling between ferroelectric and ferromagnetic degrees of freedom have attracted intense research effort due to their application potential in tunable multifunctional devices \cite{Kimura, Hur, Fiebig, Tokura1, Dong}. So-called spin-driven ferroelectrics, for which the inversion symmetry is broken in the ferroelectric phase due to the appearance of a particular magnetically ordered state, provide a path to the required coupling \cite{Cheong,Tokura2}. It was found that spin-driven ferroelectricity can occur in different magnetic materials with different types of magnetic ordering states. In order to understand magnetoelectric (ME) coupling in spin driven ferroelectrics, three different microscopic models, namely exchange striction model, inverse Dzyaloshinskii-Moriya (DM) model and spin-dependent \emph{p}-\emph{d} hybridization model, have been proposed to describe the observed ferroelectricity in different spin-driven ferroelectrics \cite{Katsura1,Mostovoy,Jia}.

The inequivalent exchange striction induced by the symmetry spin exchange interaction between the neighboring spins is considered as the driving force of ferroelectricity for multiferroics with commensurate spin order and low symmetry on the specific chemical lattice, such as Ca$_3$(Co,Mn)O$_6$ \cite{Choi} and \emph{R}Mn$_2$O$_5$ (\emph{R} = Tb-Lu) \cite{Chapon}. The inverse DM model arising from the antisymmetric spin exchange interaction between canted spin sites can be used to describe the emergence of ferroelectricity for multiferroics with transverse screw spin configurations, such as \emph{R}MnO$_3$ (\emph{R} = Gd, Tb, Dy) \cite{Goto} and Ni$_3$V$_2$O$_8$ \cite{Lawes} with cycloidal spin order, CoCr$_2$O$_4$ \cite{Yamasaki} and (Sr,Ba)$_3$Co$_2$Fe$_{24}$O$_{41}$ with transverse-conical spin order \cite{Kitagawa}. According to inverse DM model, electric polarization is produced through the spin-orbit interaction by displacing anions between two canted magnetic moments. It is worthwhile to note that multiferroicity around room temperature have been discovered in conical hexaferrite systems \cite{Kitagawa,Tokunaga1}, although the ME effect in them is still too small for practical application. Unlike exchange striction model and inverse DM model, spin-dependent \emph{p}-\emph{d} hybridization model predicts the appearance of electric polarization along certain bond direction due to the spin-direction dependent hybridization arising from the spin-orbit coupling. A typical multiferroic system based on the spin-dependent \emph{p}-\emph{d} hybridization is CuFeO$_2$ with proper screw spin order \cite{Kimura2}.

Given that strong coupling between the magnetic and ferroelectric orders exists intrinsically in spin-driven ferroelectrics, magnetic field control of spontaneous polarization and/or electric field control of the magnetization have been successfully demonstrated in several spin-driven multiferroics \cite{Yamasaki, Tokunaga2}. Moreover, the cross-control between electric and magnetic dipoles in multiferroics is found to be accompanied by the dynamical motion of multiferroic domain walls \cite{ Tokunaga2, Kagawa,Seki1}. In addition, the dynamic magnetoelectric coupling in multiferroics will lead to the appearance of electric-dipole-active magnetic resonance, i.e. a novel collective excitation named electromagnon \cite{Katsura2}. The identification of an electromagnon by terahertz spectroscopy and neutron scattering confirmed the existence of this dynamical magnetoelectric coupling \cite{Katsura2,Pimenov1, Takahashi, Senff,Jones,Finger1}.

As a prototypical multiferroic material with spiral magnetic order, MnWO$_4$ has been widely studied concerning its magnetic and ferroelectric properties \cite{Heyer, Taniguchi1, Arkenbout, Chaudhury1, Ye1, Chaudhury2, Ye2, Sagayama, Finger2,Niermann, Kumar}. It is well accepted that the inverse DM mechanism is relevant for modeling the ME coupling in MnWO$_4$. Similar to the case of multiferroic TbMnO$_3$, the magnetically induced electric polarization in MnWO$_4$ fulfills the relation:
\begin{equation}
{\vec{\textbf{P}}} = A \vec{\textbf{e}}_{ij}  \times (\vec{\textbf{S}}_i \times \vec{\textbf{S}}_j)
\end{equation}
Here \emph{A} is a coupling coefficient and $\vec{\textbf{e}}_{ij}$ is the unit vector connecting two neighboring Mn moments $\vec{\textbf{S}}_i$ and $\vec{\textbf{S}}_j$. However, it is argued that DM interaction is not the only driving force for the ferroelectric polarization, other single-site symmetric interactions are also shown to be involved in the magnetoelectric process in MnWO$_4$ \cite{Toledano}. Besides, a theoretical study suggests a more complex scenario where multiferroicity in MnWO$_4$ is caused by a competition of DM and isotropic exchange interactions \cite{Solovyev}. A deeper insight into the coupling between the electric and magnetic degrees of freedom can be gained by studying not only the respective order but also the excitation spectra.

In this paper, we report the observation of elementary magnetic excitations in multiferroic MnWO$_4$ by high resolution inelastic neutron scattering. A detailed analysis of the low-energy excitations of MnWO$_4$ shows that the spin-wave excitations in the collinear antiferromagnetic/paraelectric phase can be well described by a Heisenberg model with magnetic exchange couplings extending to the 12$^{th}$ nearest neighbor, although MnWO$_4$ was considered as a moderately spin frustrated system. In addition, the electromagnon excitation observed in both the paraelectric and ferroelectric phases supports the existence of strong spin-orbit interaction in MnWO$_4$, despite the fact that Mn$^{2+}$ is an \emph{S}-state ion.

\section{\label{sec:level1} Experiment}

Single crystals of MnWO$_{4}$ were grown by floating zone method. The heat capacity measurement was performed by using a quantum design physical property measurement system (PPMS). Single-crystal neutron diffraction and inelastic neutron spectroscopy were performed on the cold-neutron triple-axis spectrometer PANDA operated by JCNS at the MLZ in Garching, Germany \cite{Schneidewind}. The crystal used for the neutron scattering measurements has the shape of a cylinder with a total mass of 5.1 g. For the inelastic neutron scattering measurements,  pyrolytic graphite PG(002) were selected as monochromator and analyzer, while a Be filter was placed before analyzer. The final neutron wave vector was fixed to $k_f$ = 1.2 $\buildrel_\circ \over {\mathrm{A}}$$^{-1}$. In what follows we will describe the neutron scattering data in the high symmetry monoclinic \emph{P2/c} space group with scattering vector \textbf{Q} = (\emph{q$_x$}, \emph{q$_y$}, \emph{q$_z$}) (in units of $\buildrel_\circ \over {\mathrm{A}}$$^{-1}$) at position (\emph{HKL}) = (\emph{q$_x$a}/2$\pi$, \emph{q$_y$b}/2$\pi$, \emph{q$_z$c}/2$\pi$) in reciprocal lattice units, where \emph{a} = 4.8226(3) $\buildrel_\circ \over {\mathrm{A}}$, \emph{b} = 5.7533(6) $\buildrel_\circ \over {\mathrm{A}}$, \emph{c} = 4.9923(5) $\buildrel_\circ \over {\mathrm{A}}$ and  $\beta$ = 91.075(7) $\textordmasculine$ at \emph{T} = 1.5 K \cite{Lautenschl}. The MnWO$_4$ single crystal was aligned in the scattering plane defined by the orthogonal vectors (1 0 2) and (0 1 0), in which low-energy excitations along main symmetry directions in the magnetic Brillouin zone can be surveyed.

\section{\label{sec:level1} Experimental results and modeling}

\subsection{\label{sec:level2} Antiferromagnetic order in MnWO$_4$}

\begin{figure}
\includegraphics[width=8.5cm,height=9.5cm]{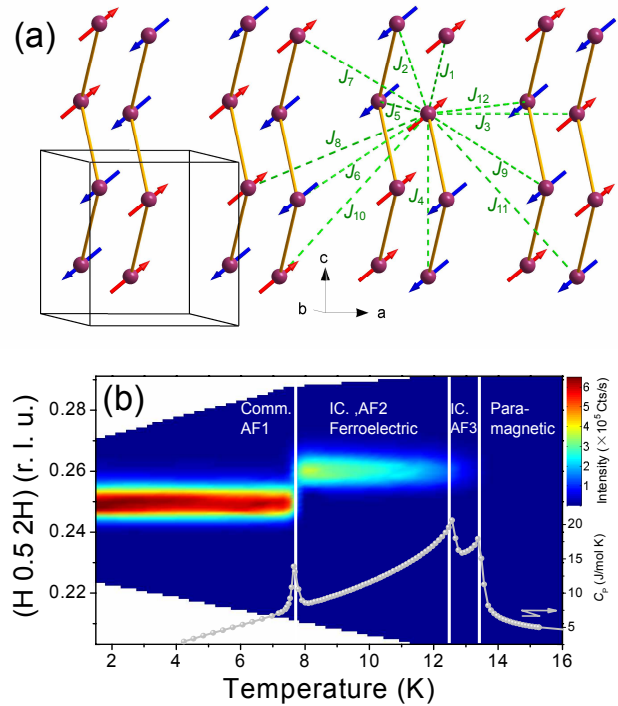}
\caption{\label{fig:epsart} (Color online) (a) Schematic drawing of the crystal and magnetic structure of the AF1 phase of MnWO$_4$ showing the magnetic Mn$^{2+}$ ions only. Solid lines highlight the monoclinic unit cell. The arrows on the Mn atoms denote the directions of their spins and the dashed lines denote the paths of Heisenberg exchange interactions.  (b) Evolution of magnetic modulation (1/4, 1/2, 1/2) with increasing temperature. The dots denotes the temperature dependence of the molar specific heat measurement. Four distinct phases labeled as  AF1, AF2, AF3 and paramagnetic are revealed. Comm. and IC. denote commensurate and incommensurate modulation, respectively.}
\end{figure}

MnWO$_4$ crystalizes in monoclinic \emph{P2/c} structure. It consists of MnO$_6$ octahedra with Mn$^{2+}$ (\emph{d}$^5$) ions as well as W$^{6+}$ (\emph{d}$^0$) ions \cite{Sleight}. High-spin Mn$^{2+}$ ion with the \emph{d}-orbital configuration \emph{t}$_{2g}^{3}$\emph{e}$_{g}^{2}$ is the only magnetic ion in the MnWO$_4$ unit cell. Below the magnetic ordering temperature (T$_N$ = 13.5 K), MnWO$_4$ undergoes three successive magnetic transitions. The corresponding phases are labeled as AF1, AF2 and AF3 \cite{Lautenschl}. The AF1 phase below 7.8 K is a commensurate antiferromagnetic phase with wave vector \emph{\textbf{k}}$_C$ = ($\pm$1/4, 1/2, 1/2) as illustrated in Fig. 1(a). Between 7.8 and 12.6 K, the AF2 phase with incommensurate spiral spin structure prevails. An incommensurate propagation vector \emph{\textbf{k}}$_{IC}$ = (-0.214, 1/2, 0.457) has been reported for the AF2 phases. With further increase in temperature the AF3 phase appears, where the Mn$^{2+}$ moment order in an incommensurate collinear antiferromagnetic configuration with the same magnetic wave vector as in AF2 phase\cite{Lautenschl}.

As shown in Fig. 1(b), the magnetic phase transitions in MnWO$_4$ are visible in both \emph{Q}-scan through the magnetic (1/4, 1/2, 1/2) reflection and specific heat curve. The lock-in transition between AF2 and AF1 takes place at 7.8 K. Subsequently, an incommensurate wave vector is observed around (0.26(1), 1/2, 0.52(1)) for the AF2 and AF3 phases. It should be noted that the observed wave vector located at (0.26(1), 1/2, 0.52(1)) in AF2 phase is equivalent to the \emph{\textbf{k}}$_{IC}$ = (-0.214, 1/2, 0.457) reported in \cite{Lautenschl}. Indeed, the signal at (0.26(1), 1/2, 0.52(1)) comes from the adjacent IC magnetic reflection (0.214, 1/2, 0.543), which can be indexed by (0 1 1)$_N$-\emph{\textbf{k}}$_{IC}$.

\subsection{\label{sec:level2} Spin-wave dispersion and analysis}

The observed low-energy excitations in the AF1 phase at 1.5 K along the [1 0 2] and [0 1 0] directions through the magnetic peak (1/4, 1/2, 1/2) are shown in Fig. 2(a) and (c), respectively. Because MnWO$_4$ is an antiferromagnetic insulator with a rather large ordered Mn moment, we analyze inelastic neutron scattering data in the linear spin-wave approximation with a Heisenberg Hamiltonian \cite{Brueckel}:
\begin{equation}
H = -\frac{1}{2}  \sum_{i,j}  J_{ij} \textbf{S}_i \cdot \textbf{S}_j -D_s \sum_{i} \textbf{S}^{2}_{i,z}
\end{equation}
\noindent Here \emph{J}$_{ij}$ denote exchange constants (from \emph{J}$_1$ to \emph{J}$_{12}$), while \emph{D}$_s$ is the uniaxial single-ion anisotropy constant.

\begin{figure}
\includegraphics[width=8.5cm,height=9cm]{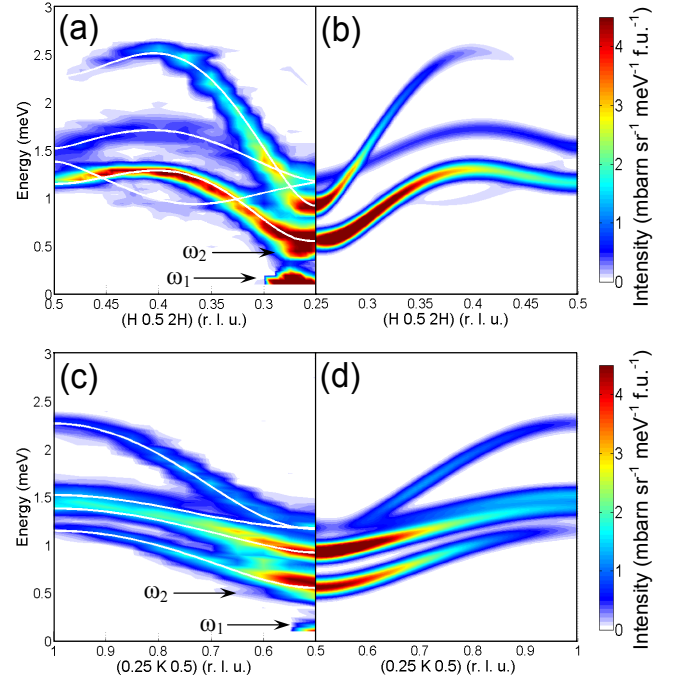}
\caption{\label{fig:epsart} (Color online) The measured spin-wave spectrum compared to calculations based on a Heisenberg model that describes the antiferromagnetic ground state of MnWO$_4$. (a) and (c) Experimental low-energy excitations along [1 0 2] and [0 1 0] directions at temperature of 1.5 K. The spectrum is composed of 16 constant-\emph{Q} scans at various wave vector. The solid lines demonstrated the spin-wave dispersion relationship resulting from a fit of the dispersion relations employing a Heisenberg Hamiltonian as described in the text. (b) and (d) Calculated spin-wave excitations of MnWO$_4$ in comparison with measured ones. The color code represents the intensity of inelastic scattered neutrons}
\end{figure}

In the collinear AF1 phase, the magnetic unit cell of MnWO$_4$ contains in total eight spins, i.e. four up spins (labeled as $i$ with $i$ = 1, 2, 3, 4) and four down spins (labeled as $j$ with $j$ = 1, 2, 3, 4).
The linearized Holstein-Primakoff transformation for the quantum spin $\hat{S}$ at each site is given as
\begin{subequations}
\begin{eqnarray}
\hat{S}_i^\dag &=&  \sqrt{2S} (1-\frac{a_i^\dag a_i}{2S})^{1/2} a_i \thickapprox \sqrt{2S} a_i, \label{appa}
\\
\hat{S}_i^- &=&  \sqrt{2S} (1-\frac{a_i^\dag a_i}{2S})^{1/2} a_i^\dag \thickapprox \sqrt{2S} a_i^\dag, \label{appb}
\\
\hat{S}_i^z &=& S - a_i^\dag a_i, \label{appc}
\end{eqnarray}
\end{subequations}
\begin{subequations}
\begin{eqnarray}
\hat{S}_{j}^\dag &=&  \sqrt{2S} (1-\frac{b_j^\dag b_j}{2S})^{1/2} b_j^\dag \thickapprox \sqrt{2S} b_j^\dag, \label{appa}
\\
\hat{S}_j^- &=&  \sqrt{2S} (1-\frac{b_j^\dag b_j}{2S})^{1/2} b_j \thickapprox \sqrt{2S} b_j, \label{appb}
\\
\hat{S}_j^z &=& -S + b_j^\dag b_j, \label{appc}
\end{eqnarray}
\end{subequations}

The Fourier transformation for bosonic operators is
\begin{subequations}
\begin{eqnarray}
\hat{a}_{i}^\dag &=&  \sqrt{\frac{1}{N}} \sum_{q} e^{-iq\cdot r_i} a_q^\dag, \label{appa}
\\
\hat{b}_{j}^\dag &=&  \sqrt{\frac{1}{N}} \sum_{q} e^{-iq\cdot r_j} b_q^\dag, \label{appa}
\\
\hat{a}_{i} &=&  \sqrt{\frac{1}{N}} \sum_{q} e^{iq\cdot r_i} a_q, \label{appa}
\\
\hat{b}_{j} &=&  \sqrt{\frac{1}{N}} \sum_{q} e^{iq\cdot r_j} b_q, \label{appa}
\end{eqnarray}
\end{subequations}

By using Holstein-Primakoff transformation, we can obtain the bosonic Hamiltonian in the momentum space as
\begin{equation}
H = - \frac{1}{2}  \sum_{q}  \psi_q^\dag \mathscr{H}_q \psi_q
\end{equation}
where $\psi_q^\dag = (a_{q1}^\dag \ a_{q2}^\dag \ a_{q3}^\dag \ a_{q4}^\dag \ b_{-q-1}^\dag \ b_{-q-2}^\dag \ b_{-q-3}^\dag \ b_{-q-4}^\dag)$,

\begin{equation}
\mathscr{H}_q=S \left(
  \begin{array}{cccccccc}
    H_0 &  A   &  B   &  C    &  E    &  F   & -B    &  C^*  \\
    A^* &  H_0 &  C   &  D    &  F^*  &  E   &  C^*  & -D    \\
    B^* &  C^* &  H_0 &  A^*  & -B^*  &  C   &  E    &  F^*  \\
    C^* &  D^* &  A   &  H_0  &  C    & -D^* &  F    &  E    \\
    E   &  F   & -B   &  C^*  &  H_0  &  A   &  B    &  C    \\
    F^* &  E   &  C^* & -D    &  A^*  &  H_0 &  C    &  D    \\
    -B^*&  C   &  E   &  F^*  &  B^*  &  C^* &  H_0  &  A^*  \\
    C   & -D^* &  F   &  E    &  C^*  &  D^* &  A    &  H_0    \\
    \end{array}      \nonumber
\right)
\end{equation}

with
\begin{subequations}
\begin{eqnarray}
    H_0 &=& 2(J_4+J_5+J_6-J_7-J_8+J_9-D_s)      \nonumber
    \\
    A &=& J_7e^{ iq_x+(2y-1)iq_y-0.5iq_z}+J_7e^{-iq_x+(2y-1)iq_y+0.5iq_z}       \nonumber
    \\
    &&+J_8e^{-iq_x+(2y-2)iq_y-0.5iq_z}+J_8e^{ iq_x+(2y-2)iq_y+0.5iq_z}       \nonumber
    \\
    B &=& J_1e^{ (2y-1)iq_y+0.5iq_z}+J_2e^{ (2y-2)iq_y-0.5iq_z}       \nonumber
    \\
    C &=& J_3e^{ iq_x}+J_{10}e^{-iq_x-iq_z}+J_{11}e^{-iq_x+iq_z}+J_{12}e^{-iq_x-iq_y}      \nonumber
    \\
    &&+J_{12}e^{-iq_x+iq_y}      \nonumber
    \\
    D &=& J_1e^{ (1-2y)iq_y+0.5iq_z}+J_2e^{ (2-2y)iq_y-0.5iq_z}       \nonumber
    \\
    E &=& 2J_4 cos(q_z)+2J_5 cos(-q_y)       \nonumber
    \\
    F &=& J_6e^{-iq_x+(2y-1)iq_y-0.5iq_z}+J_6e^{iq_x+(2y-1)iq_y+0.5iq_z}      \nonumber
    \\
    &&+J_9e^{iq_x+(2y-2)iq_y-0.5iq_z}+J_9e^{-iq_x+(2y-2)iq_y+0.5iq_z}      \nonumber
\end{eqnarray}
\end{subequations}

In order to obtain the eigenvalues and eigenvectors for the spin-wave modes, the bosonic Hamiltonian is diagonalized numerically. To interpret the inelastic neutron scattering data, we connect the experimental results with theoretical calculation by using inelastic neutron scattering cross section. The inelastic neutron scattering cross section can be written as:
\begin{equation}
\frac{d^2\sigma}{d\Omega d E}  = \frac{k_f}{k_i} \left( \frac{1}{2}   \gamma r_0 \text{g}   F(\textbf{Q}) \right)^2 e^{-2W}\sum_{\alpha,\beta}(\delta_{\alpha,\beta}-\hat{Q}_\alpha \hat{Q}_\beta)S^{\alpha \beta}(\textbf{Q},\omega)
\end{equation}
Here $k_f$ and $k_i$ are final and incident wave vectors, respectively. $\gamma r_0 = 5.39 fm$ is the magnetic scattering amplitude for an electron. $\text{g}$ is the Land$\acute{e}$ $\text{g}$-factor of Mn, $F(\textbf{Q})$ is the form factor for magnetic ion Mn$^{2+}$, $e^{-2W}$ is the Debye-Waller factor. $\hat{Q}_\alpha$ is the component of a unit vector in the direction of \emph{\textbf{Q}}, and $S^{\alpha \beta}(\textbf{Q},\omega)$ is the response function describing spin correlations \cite{Lovesey}. Only the transverse correlations contribute to the linear spin-wave cross section through single magnon excitations.

As mentioned above, the primitive magnetic unit cell of MnWO$_4$ is composed of eight Mn spins in the collinear AF1 phase. Therefore, four two-fold degenerate spin-wave branches are expected in zero field. However, the excitation spectrum at the zone center \textbf{Q} = (1/4, 1/2, 1/2) exhibits at least five resolvable excitations. This can be clearly seen in the individual energy-scans plotted in Fig. 5(a). By considering different possibilities and evaluating the results of the refinements as function of \textbf{Q}, we found that it is impossible to describe the two low-lying energy excitations located in the zone center at $\omega_1$ = 0.07(1) and $\omega_2$ = 0.45(1) meV within the Heisenberg model. As demonstrated in Figs. 2(b) and (d), the dispersion as well as the intensities of the magnetic excitations can be well modeled as spin-wave excitations using the Hamiltonian in Eq. (2), if the low-lying excitations $\omega_1$ and $\omega_2$ are excluded. The fitting results exhibit excellent agreement with the experimental data for the spin-wave excitations and yield exchange parameters as $J_1$ = -0.37(1), $J_2$ = -0.002(1), $J_3$ = -0.17(1), $J_4$ = -0.21(1), $J_5$ = -0.011(5), $J_6$ = -0.34(1), $J_7$ = -0.11(1), $J_8$ = -0.010(5), $J_9$ = -0.20(1), $J_{10}$ = -0.12(1), $J_{11}$ = -0.042(1), $J_{12}$ = -0.016(1) meV, and $D_S$ = 0.06(1), where all parameters are given in units of meV. All obtained exchange parameters are negative, which indicates that three dimensional antiferromagnetic exchange interaction is the dominant interaction in MnWO$_4$. It is also noticed that the strengths of some exchange interactions, such as $J_3$ and $J_6$, are comparable to that of the nearest neighbor interaction $J_1$, suggesting the existence of substantial magnetic frustration in MnWO$_4$.

\begin{figure}
\includegraphics[width=8.5cm,height=6.5cm]{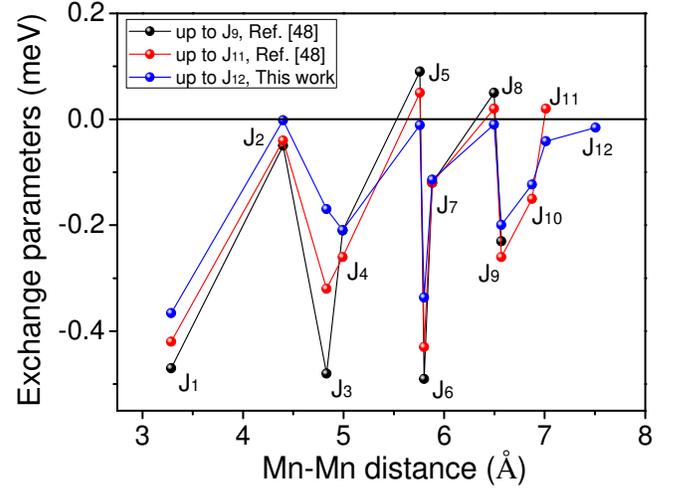}
\caption{\label{fig:epsart} (Color online) Distance dependence of exchange interaction parameters. The parameters are obtained by considering Mn-Mn spin exchange interaction up to 9$^{th}$, 11$^{th}$ and 12$^{th}$ nearest neighbor, respectively. }
\end{figure}

The spin-wave excitations in AF1 phase of MnWO$_4$ have been studied previously by several groups, and the exchange parameters have been extracted from neutron scattering experiments as well as from theory \cite{Ehrenberg, Tian, Ye3}. However, the parameters deduced in previous studies failed to reproduce the dispersion spectra in some aspects. In latest reference \cite{Ye3}, Ye \emph{et al.} firstly modeled experimental results by considering nine exchange parameters at the beginning, but the authors found that more satisfied agreement can be achieved by considering exchange parameters up to $J_{11}$. As shown in Fig. 3, non-negligible values was obtained for $J_{10}$ and $J_{11}$ accompanied with the weakening or strengthening of other exchange interactions. However, it can be seen that the Heisenberg model including up to 11$^{th}$ exchange parameters is still not enough to fit the observed neutron spectra properly. For instance, the dispersion spectra along [1 0 -2] direction through the magnetic peak (1/4, 1/2, -1/2), especially the low lying branch, cannot be reproduced properly, as indicated in Fig. 3 of Ref.\cite{Ye3}. In the present work, we found that the fitting of the spectra can be further improved if we include one more exchange parameter $J_{12}$. It is worth noting that the number of neighbors is two for exchange couplings $J_{1}$, $J_{2}$,.., and $J_{11}$, whereas it is four for exchange coupling $J_{12}$, indicating the considerable weight of exchange interaction between one given spin and its 12$^{th}$ nearest neighbor. The variation of exchange interaction parameters in the present work as a function of Mn-Mn distance is also plotted in Fig. 3 for comparison. It can be seen that most of the exchange parameters changed monotonically if more exchange parameters are took into account. The obtained parameters can provide excellent agreement between theoretical and experimental results throughout the entire Brillouin zone. The calculated spin-wave excitation spectra along both [1 0 2] and [1 0 -2] directions are plotted in Fig. 4. Compared with experimental and calculated neutron spectra shown in Ref.\cite{Ye3}, the calculated spectra in the present work show significantly better agreement with the experimental data.

\begin{figure}
\includegraphics[width=8.5cm,height=6.5cm]{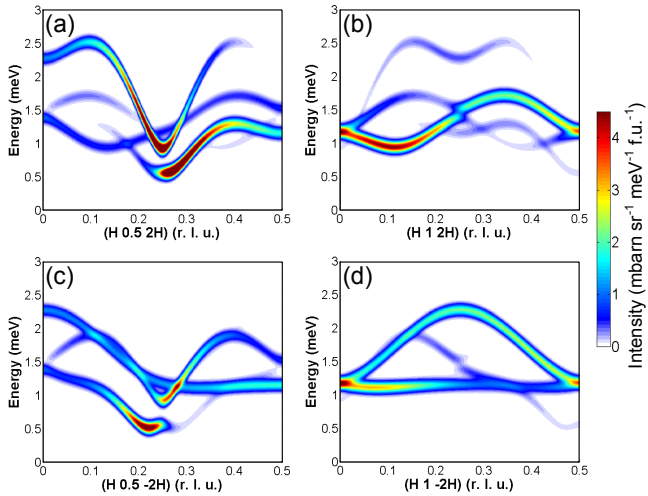}
\caption{\label{fig:epsart} (Color online) (a) and (b) Calculated spin-wave excitation spectra of AF1 phase along [1 0 2] direction with K = 0.5 and 1, respectively. (c) and (d) Calculated spin-wave excitation spectra of AF1 phase along [1 0 -2] direction with K = 0.5 and 1, respectively. }
\end{figure}

\subsection{\label{sec:level2} Electromagnon excitation and dispersion}

\begin{figure}
\includegraphics[width=8.5cm,height=12.5cm]{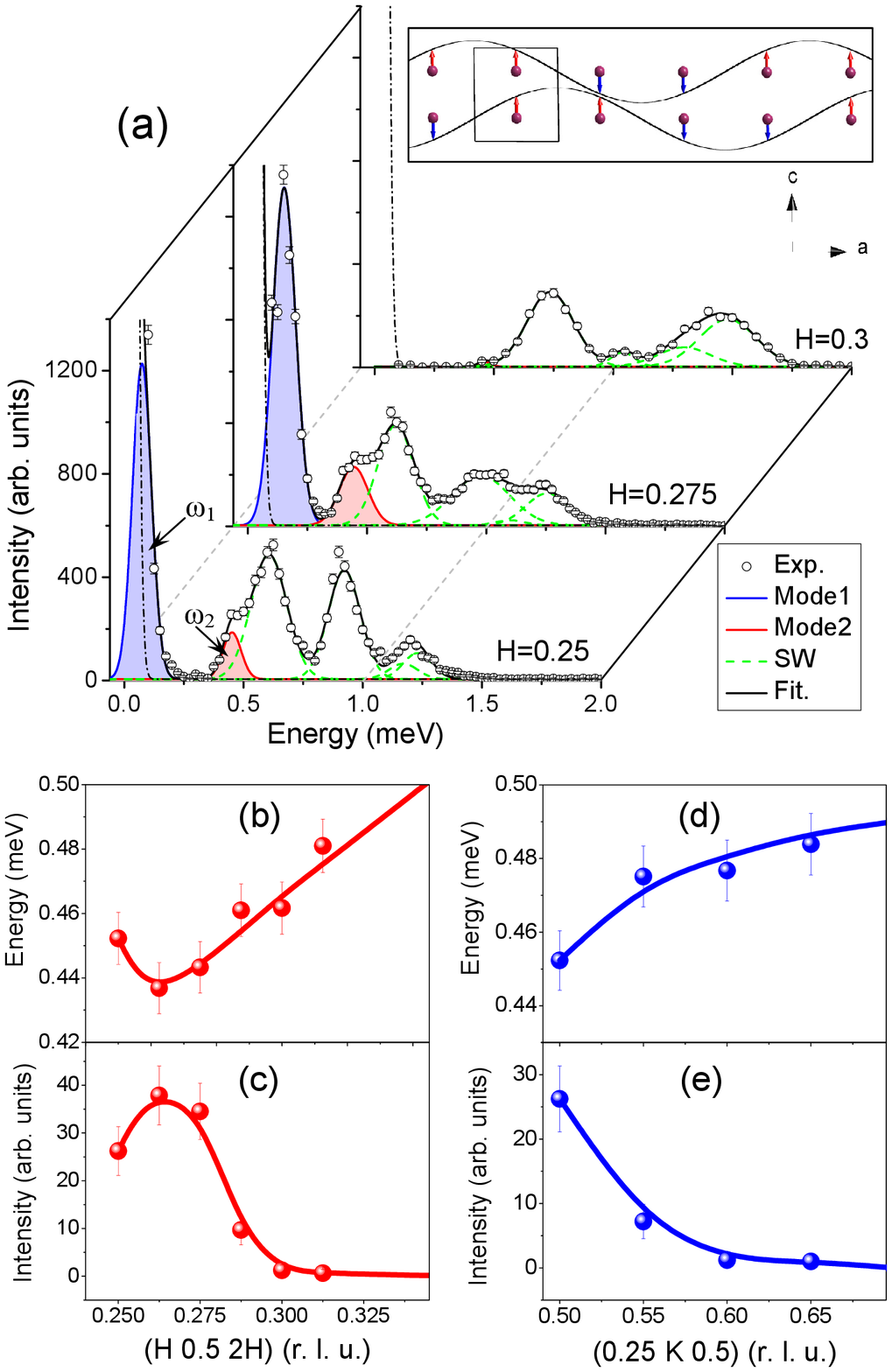}
\caption{\label{fig:epsart} (Color online) (a) Representative magnetic excitation spectra around zone center (1/4, 1/2, 1/2) at 1.5 K. The dashed lines are the estimated line shape for spin-wave excitations, while the spectral weights arising from electromagnon excitations are highlighted as the shaded areas under the curves. The dash-dotted line corresponds to the instrumental resolution. The inset illustrated the $\uparrow\uparrow\downarrow\downarrow$ spin configuration along the \emph{a}-axis, which can be treated as a special case of a modulated spin structure. (b) and (c) Dispersion relation and spectral weight of electromagnon excitation $\omega_2$ along [1 0 2] direction. The solid lines are guides to the eye. (d) and (e) Dispersion relation and spectral weight of electromagnon excitation $\omega_2$ along [0 1 0] direction.}
\end{figure}

Although the ME coupling in Multiferroics will lead to the emergence of electromagnon, the electromagnon excitation is not restricted to multiferroics since the macroscopic origin of dynamical ME coupling does not necessarily produce the multiferroic ground state. For instance, electromagnon excitations are observed in paramagnetic phase of TbMnO$_3$ \cite{Senff2} and CuFe$_{1-x}$Ga$_x$O$_2$ \cite{Seki2} with collinear spin structure. In the AF1 phase of MnWO$_4$, the Mn$^{2+}$ spins are aligned collinearly along the easy axis with the spin direction alternating along the \emph{a}-axis as $\uparrow\uparrow\downarrow\downarrow$ (see inset of Fig. 5(a)). The $\uparrow\uparrow\downarrow\downarrow$ spin configuration can be considered as the special case of a modulated structure in form of
$\textbf{S}_i = \textbf{S}_0 \cdot $cos$(2\pi k R_1 + \pi/4)$ with \emph{k} = 1/4. As for a spiral magnet, one phason mode and two rotation modes may contribute to the low-energy excitation. The observed two modes $\omega_1$ and $\omega_2$ in Fig. 5(a) can be attributed to the electromagnon excitations and they relate to the phason and rotation modes, respectively. In Figs. 5(b) and (c), the dispersion relation along the [1 0 2] direction as well as the intensity change of electromagnon mode $\omega_2$ extracted from the individual energy scans are plotted. Interestingly, an energy dip in the dispersion relation and a peak in the intensity are observed at \textbf{Q} = (0.26(1), 1/2, 0.52(2)), which corresponds exactly to the magnetic propagation vector we find for the incommensurate AF2 phase, see Fig. 1(b). The minimal energy gap for the electromagnon  $\omega_2$ is associated with the magnetoelectric coupling effect in multiferroic MnWO$_4$. In contrast to the dispersion along the [1 0 2] direction, both energy and intensity of the electromagnon $\omega_2$ along [0 1 0] direction evolute monotonically away from \textbf{Q} = (1/4 1/2 1/2), as shown in  Fig. 5(d) and (e), indicating an anisotropic dispersion behaviors of electromagnons in MnWO$_4$.

\begin{figure}
\includegraphics[width=8.5cm,height=6cm]{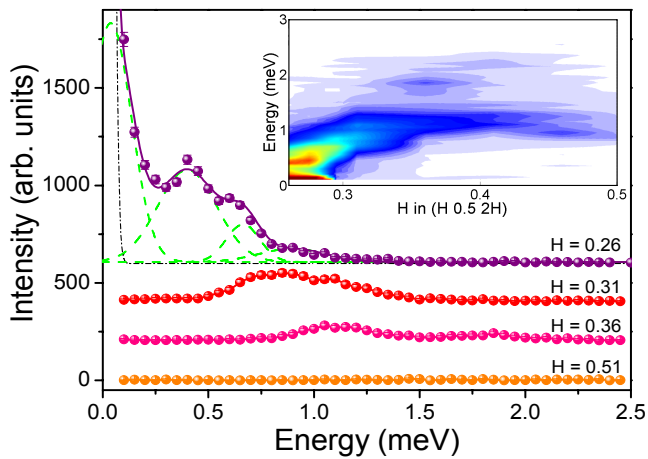}
\caption{\label{fig:epsart} (Color online) (a) Selected energy-scans at various wave vector in the incommensurate AF2 phase at 10 K. The curves are shifted vertically for comparison. The inset shows the measured excitations in the AF2 phase around the zone center \textbf{Q} = (0.26, 0.5, 0.52) as a color plot.}
\end{figure}

Upon the increase in temperature, a transition from the collinear phase AF1 into the ferroelectric magnetic spiral phase AF2 occurs. In Fig. 6, representative scans of the low-energy excitations and their dispersion in the AF2 phase at 10 K are shown. Although the dispersion behavior persists in momentum space, the excitations broaden and soften in the AF2 phase in comparison to the AF1 phase. Nevertheless, four well resolved excitation modes are identified by fitting the energy scans at the incommensurate wave vector \textbf{Q} = (0.26, 0.5, 0.52). Among these four modes, the broad mode located at 0.04(2) meV most likely corresponds to the $\omega_1$ mode in the AF1 phase. As one of the electromagnon excitation modes, the $\omega_1$ mode in the ferroelectric AF2 phase shows substantial spectrum weight gain about two times compared to the non-ferroelectric AF1 phase. This underpins our identification of the low-energy excitation mode $\omega_1$ as being associated with the dynamical magnetoelectric coupling.

Since electromagnon excitation is an electrically active spin excitation related to the multiferroic character in multiferroics, many efforts have been made aiming to elucidate the feature of electromagnon in spin-driven multiferroics. The infrared spectroscopy and polarized neutron spectroscopy were accepted to be effective methods in evidencing the existence of electromagnon  \cite{Pimenov1,Senff}. In Ref.\cite{Choi}, the authors carried out optical spectroscopic investigations on MnWO$_4$, but no signal from electromagnon was found in the low-energy region of optical spectra with energy down to 0.62 meV, indicating that the electromagnon excitation in MnWO$_4$ might locate at a lower energy region. As a matter of fact, in the present work the observed two modes $\omega_1$ and $\omega_2$ are all below 0.6 meV. Nevertheless, it is impossible for us to identify the character of the low-lying energy modes from unpolarized neutron spectroscopy results. Further experimental work via polarized neutron spectroscopy is necessary in order to identify the observed magnon modes, thus to firmly confirm that the low-lying energy excitations are electromagnon excitations in nature.

\section{\label{sec:level1}  Discussion}

From our analysis of the spin-wave dispersion in the framework of the Heisenberg model, we deduced a spin-wave energy gap $\Delta$ of 0.61(1) meV and a single-ion anisotropy constant \emph{D}$_s$ of 0.06(1) meV for the AF1 phase. The observed anisotropy is considerable given the fact that in a purely ionic description, manganese in MnWO$_4$ is an \emph{S}-state Mn$^{2+}$  ion with vanishing orbital moment. Full multiplet calculation shows that the magnetic anisotropy in MnWO$_4$ arises from the spin-orbit coupling and it is comparable in energy to the Dzyaloshinskii-Moriya interactions \cite{Hollmann}.

Compared to the spin-wave energy gap, the energy gaps of 0.07(1) meV and 0.45(1) meV for the two observed electromagnon excitations are relatively small, but of the same order of amplitude. If the inverse DM interaction is considered as the mechanism responsible for the multiferroic properties in MnWO$_4$, the observed electromagnon excitations, which are associated with the magnetoelectric coupling, may arise from the DM exchange interaction instead of the Heisenberg interaction. Moreover, although the two observed electromagnon excitation modes cannot be described within the Heisenberg model, they do exhibit dispersive behavior characteristic for collective excitations. As shown in Fig. 5, the energy of $\omega_2$ decreases slightly between \textbf{Q} = (1/4, 1/2, 1/2) and \textbf{Q} = (0.275, 1/2, 0.55), whereas the energy of $\omega_1$ increase from 0.07(1) meV to 0.15(1) meV. The opposite dispersion behavior of the two electromagnon modes around the zone center suggests that the origins of the two excitation modes are different and/or these two excitation modes might couple with each other through hybridization effects. Further experimental works via complementary techniques and comprehensive dynamical structure calculations are required to get further insight into the nature of electromagnon excitations. It is also known that the application of magnetic field along the easy-axis of MnWO$_4$ will lead to a switch of the polarization accompanied with the magnetic phase transition \cite{Taniguchi2}. Therefore, it is interesting to investigate the evolution of electromagnon excitations upon the application of magnetic field in MnWO$_4$. The behavior of electromagnon excitations across the field-induced transition between ferroelectric and paraelectric phases can be helpful to identify the character of electromagnon excitations \cite{Pimenov2,Holbein}, thus leading to a better understanding of the mechanism of magnetoelectric coupling in multiferroics with noncollinear spiral magnetic structure.

\section{\label{sec:level1} Conclusion}

In summary, we present results of a comprehensive neutron scattering study of the elementary magnetic excitations in multiferroic MnWO$_4$. In addition to the well-known spin-wave excitations, we demonstrate the existence of electromagnon excitations in both the paraelectric AF1 and the ferroelectric AF2 phases. The spin-wave excitations of MnWO$_4$ in the AF1 phase can be properly described by a Heisenberg model with local magnetic exchange coupling extending up to the 12$^{th}$ nearest neighbor. The analysis of spin-wave excitations suggests that the dominant interaction are antiferromagnetic in MnWO$_4$. Competing antiferromagnetic exchange leads to frustration and gives rise to the modulated magnetic phases and the rich phase diagram. The spin-wave gap of 0.61(1) meV at 1.5 K amounts to roughly 1/4 of the zone boundary spin-wave energy. This relatively large value indicates the existence of rather strong spin-orbit interaction, despite the fact that Mn$^{2+}$ is an \emph{S}-state ion. The analysis of the low-energy excitation spectra implies the existence of collective electromagnon excitations, which reflect the strong ME coupling. Similar to the spin-wave excitations, the electromagnons also exhibit dispersive behavior with smaller but considerable energy gaps at the zone center of 0.07(1) meV and 0.45(1) meV at 1.5 K, respectively. These modes persist in both, the collinear magnetic and paraelectric AF1 phase below 7.8 K and the spin spiral ferroelectric AF2 phase between 7.8 K and 12.6 K. Taking into account the assumed mechanism for multiferroicity in MnWO$_4$ based on the inverse DM effect, we argue that the electromagnons are associated with the magnetoelectric coupling arising from the Dzyaloshinskii-Moriya exchange mechanism, which is allowed in MnWO$_4$ due to the low symmetry and the rather strong spin-orbit coupling already evidenced by the spin-wave dispersion.

\section{\label{sec:level1}Acknowledgment}

The authors are grateful to J. Per{\ss}on for providing assistance with single crystal growth and alignment. Y.X. acknowledges I.V. Solovyev for helpful discussions.

\appendix

\end{document}